\documentclass[twocolumn]{aastex62}
\usepackage{amsmath,amstext}
\usepackage[T1]{fontenc}

\shorttitle{migration timescales and Kuiper belt inclinations}

\begin{document}

\title{Not a simple relationship between Neptune's migration speed and Kuiper belt inclination excitation} 

\author[0000-0001-8736-236X]{Kathryn Volk}
\correspondingauthor{Kathryn Volk}
\email{kvolk@lpl.arizona.edu}
\affiliation{Lunar and Planetary Laboratory, The University of Arizona, 1629 E University Blvd, Tucson, AZ 85721}

\author[0000-0002-1226-3305]{Renu Malhotra}
\affil{Lunar and Planetary Laboratory, The University of Arizona, 1629 E University Blvd, Tucson, AZ 85721}

\begin{abstract}
We present numerical simulations of giant planet migration in our solar system and examine how the speed of planetary migration affects inclinations in the resulting population of small bodies (test particles) scattered outward and subsequently captured into Neptune's 3:2 mean motion resonance (the Plutinos) as well as the hot classical Kuiper belt population. 
We do not find a consistent relationship between the degree of test particle inclination excitation and e-folding planet migration timescales in the range $5-50$~Myr. 
Our results present a counter-example to \citet{Nesvorny:2015}'s finding that the Plutino and hot classical inclinations showed a marked increase with increasing e-folding timescales for Neptune's migration. 
We argue that these differing results are likely due to differing secular architectures of the giant planets during and after migration. 
Small changes in the planets' initial conditions and differences in the numerical implementation of planet migration can result in different amplitudes of the planets' inclination secular modes, and this can lead to different final inclination distributions for test particles in the simulations.
We conclude that the observed large inclination dispersion of Kuiper belt objects does not require Neptune's migration to be slow;
planetary migration with e-folding timescales of 5, 10, 30, and 50 Myr can all yield inclination dispersions similar to the observed Plutino and hot classical populations, with no correlation between the degree of inclination excitation and migration speed.
\end{abstract}

\keywords{Kuiper belt: general}

\section{Introduction}\label{s:intro}

The orbital planes of Kuiper belt objects (KBOs) are widely dispersed, and dynamical sub-classes of KBOs have measurably different inclination distributions~\citep[see, e.g.,][]{Gulbis:2010,Petit:2011,Gladman:2012,Petit:2017}. 
The large inclinations of some observed Kuiper belt objects have prompted several theoretical studies of inclination excitation during planetary migration. 
\citet{Malhotra:1995} suggested that slower migration of Neptune correlated with higher inclination excitation due to argument-of-perihelion libration within mean motion resonances.
\citet{Gomes:2003} detailed how a high-inclination population could be produced during migration as scattered KBOs were temporarily captured into Neptune's mean motion resonances, their inclinations excited by the secular Kozai-Lidov cycles, then became resonance drop-outs during the low-eccentricity/high-inclination phase of the Kozai-Lidov cycle.
Inclination excitation of KBOs has also been examined in the context of a giant planet instability during which Neptune may have been scattered out close to its current semi-major axis but on a somewhat eccentric orbit and subsequently experienced only a short-distance migration and eccentricity damping \citep[e.g.,][]{Tsiganis:2005}. 
\citet{Levison:2008} found that some simulations of such giant planet instability produced large inclinations in the resulting Kuiper belt population while others did not.

In a recent numerical study, \citet{Nesvorny:2015} found that slower migration timescales for Neptune led to larger inclinations in the hot classical population and in the Plutino population (objects in Neptune's 3:2 mean motion resonance), concluding models in which Neptune migrated from an initial semimajor axis of $a_{N,0}\lesssim25$~au to its current orbit on timescales $\tau_a~\gtrsim~10$~Myr provided the best match to observed inclination distributions. 
In this scenario, the present-day hot classical and resonant populations consist of objects originating in dynamically cold orbits interior to Neptune's current semi-major axis ($\sim~30$~au) that were scattered and dispersed by the outward migration of Neptune. 
During this outward migration, the scattered objects have a propensity to stick to Neptune's migrating mean motion resonances and undergo secular cycling of eccentricity and inclination within those resonances, as described by \citet{Gomes:2003}; most are lost from the Kuiper belt during this process, but some end up being implanted into long-term stable orbits in the hot classical region or in Neptune's exterior mean motion resonances.
\citet{Nesvorny:2015} argued that slower migration timescales allowed more time for both encounters with Neptune and secular effects within mean motion resonances to excite particle inclinations, possibly accounting for the link between migration timescale and more widely dispersed final inclinations in those simulations.

In this paper, we describe our effort to understand the dynamical mechanisms underlying the inclination excitation that occurs during planetary migration using simulations broadly similar to those of \citet{Nesvorny:2015}. 
To our surprise, we find that the degree of inclination excitation of the Plutinos and the hot classical Kuiper belt is not simply monotonically dependent on Neptune's migration timescale, but rather appears to depend sensitively on the strength of the secular inclination modes of the planets during migration (which in turn depends sensitively on the orbital evolution of the planets). 
While the simplified migration simulations presented here are not meant to mimic the full dynamical history of the outer solar system, they are useful for demonstrating that the long-timescale inclination excitation mechanisms associated with scattering and mean motion resonance sticking are not always the dominant source of inclination excitation during Neptune's migration.
We instead find that significant secular inclination excitation is possible on timescales that are short compared to typical migration timescales, especially in the vicinity of the $\nu_{18}$ secular inclination resonance.

This means that the widely dispersed inclinations in the Kuiper belt cannot be used to definitively argue for slow Neptune migration. 
An improved, more detailed understanding of how inclinations are excited during planetary migration is necessary in order to use the inclinations to constrain the history of the outer solar system. 
This is particularly important because inclination distributions are often better observationally constrained than other orbital parameters.

The rest of this paper is organized as follows.
In Section~\ref{s:sims}, we describe our simulations and results for planet migration timescales spanning an order of magnitude in e-folding timescales ($\tau_a~=~5-50$~Myr) for the planets' semi-major axis evolution. 
In Section \ref{s:discussion}, we offer an analysis of the discrepancies between our results and those of \citet{Nesvorny:2015}.
We summarize and conclude in Section~\ref{s:sum}.

\section{Numerical Simulations of Planetary Migration}\label{s:sims}

In order for planetary migration simulations to contain enough particles to sufficiently explore the orbital architecture of the resulting Kuiper belt populations, many simplifying assumptions must be made. 
In Section~\ref{ss:scheme} we describe some of the challenges of simulating planetary migration, outlining our approach to the problem and comparing that to those used in the literature. 
Section~\ref{ss:details} provides a detailed description of our numerical simulations and initial conditions.

\subsection{Background and Choice of Migration Scheme}\label{ss:scheme}

In the absence of computational limitations, one would simulate the planetesimal driven outward migration of Neptune to its current orbit by modeling the full gravitational interaction between the planets and an initial population of massive proto-Kuiper belt objects in an N-body simulation. 
In such a simulation, the semimajor axis, eccentricity and inclination evolution for Neptune (and the other planets) would occur self-consistently as the planets interact with a large number of self-gravitating objects with a realistic mass distribution.  
However, this approach is not currently feasible.
Full N-body simulations of the planet migration process are typically limited to interactions between the planets and a moderate population of non-self-interacting planetesimals. 
The planetesimal population is usually much fewer in number and thus comprised of objects that are individually more massive than expected in the conditions in the real early solar system; see, for example, the N-body work in the original Nice-model papers, where the proto-Kuiper belt is represented by $\sim~1000$ objects with individual masses ranging from a few Pluto masses to a few Lunar masses \citep{Tsiganis:2005}.
The neglect of self-gravity between the planetesimals in such populations has, at least, been shown to not have a significant impact on the final orbits of the planets \citep{Fan:2017}.
However, the planets in these simulations experience a smaller number of stronger  interactions with the planetesimals to arrive at their final orbits rather than a large number of weaker interactions, which will impact the time-evolution of their orbits. 
The computational challenges of such simulations are further compounded by their chaotic nature (although we note below that even simplified simulations suffer from this problem); a large number of sets of initial conditions must be tried before an acceptable outcome (where the giant planets end on orbits similar to their current ones) can be found \citep[as discussed in, e.g.,][]{Nesvorny:2011,Nesvorny:2012,Gomes:2018}.
Even when successful simulation initial conditions are found, there are often too few of the planetesimals surviving to the end in the simulation to have a statistically useful sample for comparison with observations of Kuiper belt objects, especially for comparisons to the dynamical sub-classes of the Kuiper belt (though for large suites of simulations, comparisons are sometimes possible, see, e.g., \citealt{Gomes:2018}).

These computational limitations mean that in simulations intended to explore the origins of the Kuiper belt dynamical structures, the planetesimal-driven migration of the planets is often instead modeled by using prescribed extra forces on the planets to cause their orbital migration, allowing the dynamical evolution of the Kuiper belt to be tracked with large numbers of massless test particles \citep[see, e.g.,][]{Hahn:2005,Levison:2008,Nesvorny:2015,Kaib:2016}.
This ``short cut" allows better statistics for the orbital distribution of a final Kuiper belt model.
The drawback of this approach is that different choices can be made as to how migration is numerically implemented, including whether and how the planets' orbital inclinations and eccentricities are damped during their migration.
For example, smooth planet migration can be modeled most simply by applying a smoothly declining torque to each planet such that its semimajor axis approaches its current observed value with the desired e-folding migration timescale; this is the approach taken in many studies \citep[e.g.][]{Malhotra:1993,Gomes:2000,Chiang:2002,Hahn:2005,Brasser:2009,Dawson:2012,Nesvorny:2015}. 
For suitable initial conditions, the planets in such simulations can maintain eccentricities and inclinations similar to their currently observed ones throughout the duration of the simulation.
In contrast, works such as \citet{Levison:2008} include eccentricity damping forces in order to model a post-instability giant planet system such as proposed in \cite{Tsiganis:2005}, wherein Neptune starts with a large eccentricity that damps down as it migrates several au to approximately its current orbit.
Other works, such as \citet{Dawson:2012} and \citet{Wolff:2012}, also choose to include eccentricity damping, but use a different numerical implementation. 
Whereas \cite{Levison:2008} implement an extra force based on expected friction in nebular gas, \citet{Wolff:2012} use analytical orbit perturbation equations to construct an extra force that yields the desired eccentricity damping behavior.
\citet{Nesvorny:2015} implemented both eccentricity and inclination damping parameterized by e-folding timescales to evolve to the planets' desired final orbits.
As we discuss further in Section~\ref{s:discussion}, it is unclear how these different choices about whether and how to implement planetary eccentricity/inclination damping affect the simulation outcomes for the test particles.

In order to limit the number of free parameters in our migration simulations, we chose to find initial conditions that avoid overly exciting the planets' eccentricities and inclinations, eliminating the need for eccentricity and inclination damping. 
As described in more detail in Section~\ref{ss:details}, we assign the planets' initial eccentricities and inclinations to be similar to their current observed values, run a large set of trial simulations, then choose the planetary initial conditions that result in satisfactory final planetary orbits. 
(We do not directly simulate scenarios in which Neptune's initial eccentricity and inclination is excited to large values by close encounters with other planets.) This is sufficient for the purposes of our study because the correlation between migration timescale and inclination dispersion found by \citet{Nesvorny:2015} did not depend on the initial assumed eccentricity or inclination for Neptune.
For simplicity, we also do not consider additional complications, such as the smoothness or graininess of Neptune's migration \citep[as investigated by, e.g.,][]{Nesvorny:2016,Kaib:2016}, which would depend on the mass distribution of the planetesimals driving the planets' migration; we simply implement smooth semimajor axis migration in the same way as \citet{Hahn:2005}. 
Our conclusion that the secular architecture of the planetary system can strongly affect the inclination distributions in the final Kuiper belt populations should hold even for grainy migration, although the efficiency of trapping test particles into the Plutino and hot classical populations could differ.

\subsection{Our Numerical Simulations}\label{ss:details}

We performed our suite of planetary migration simulations using the {\sc hermes} integrator within the {\sc rebound} software package \citep[the initial release of which is described in][]{Rein:2012}. 
This integrator utilizes a Wisdom-Holman scheme \citep{Wisdom:1991} referred to as the {\sc whfast} routine (described in \citealt{Rein:2015}) for the majority of the simulation's timesteps and switches to the {\sc ias15} adaptive stepsize routine \citep[based on][]{Everhart:1985} to resolve close encounters between objects in the simulation \citep[described in][]{Rein:2015ias}. 
At each timestep, we implement a user-defined force that causes the planets' semimajor axes to approach their current values with a prescribed e-folding timescale \citep{Hahn:2005}.
As described below, we chose initial conditions for which no close encounters occur between the planets, so their evolution is entirely calculated with {\sc whfast}. 
We chose to use {\sc rebound} (instead of one of the other hybrid numerical integration schemes in common use for solar system dynamics, such as Mercury \citep{Chambers:1999} or SWIFT \citep{Duncan:1997}) because {\sc rebound} uniquely allows for bit-wise reproducible results regardless of the choice of compiler or the operating system of the computer on which the simulations are run when the same initial inputs are used \citep[as described in][]{Rein:2017}. 
To correct the unphysical situation of the planets' evolution being affected by close encounters with massless test particles, we made a slight modification to the {\sc hermes} routine to allow bitwise reproducible histories of the planets' orbits: 
after a call to the adaptive step size routine to resolve a close encounter between a planet and a test particle, the integration for the massive bodies (in our case, the giant planets) resumes from the previous primary timestep rather than resuming from the endpoint of the close encounter; this modification amounts to discarding the planetary positions and velocities calculated during the close encounter and instead smoothly continues the integration with the values calculated from {\sc{whfast}}.   
This modification ensures that the planets' orbital histories are calculated entirely with the widely used and tested Wisdom-Holman routine and are thus unaffected by the presence or absence of test particles in the simulation.
We note that the rmvs3 routine in SWIFT requires a similar modification to ensure that interactions with massless test particles do not result in different planet outcomes.

\begin{figure*}
\centering
   \includegraphics[width=7in]{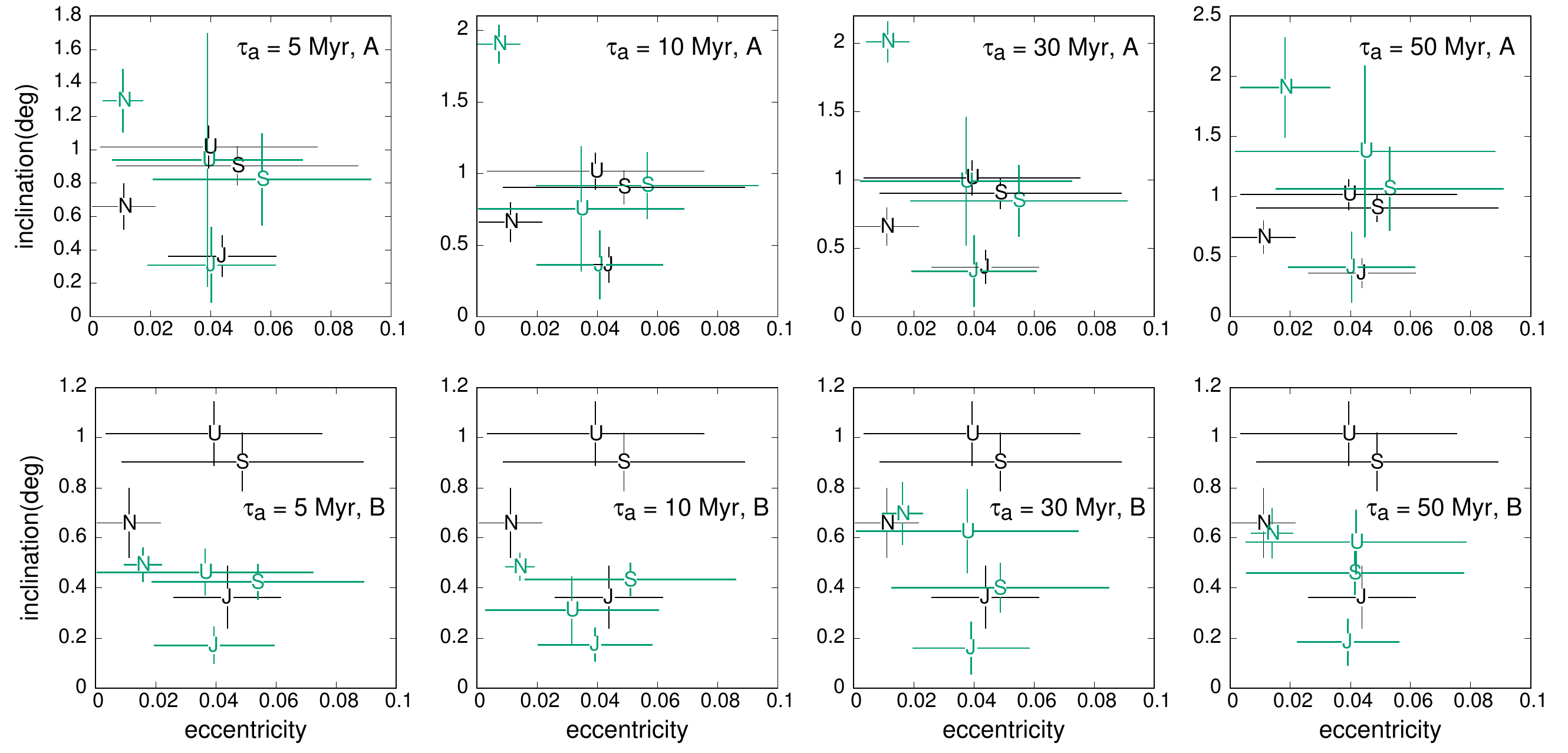} 
   \caption{
   The eccentricity and inclination ranges of the giant planets over the last 100~Myr in our migration simulations (green bars, labeled by planet) compared to those for the real solar system over the same timescale (black bars, labeled by planet). The top row of panels shows the ranges for our `A' simulations (in which Neptune's inclination is slightly larger than in the real solar system); the bottom row of panels shows the `B' simulations (in which the planets' inclinations are slightly lower than in the real solar system).  All inclinations are measured relative to the plane defined by the planets' total angular momentum vector (i.e., the invariable plane). 
   \label{f:planets}}
\end{figure*}

The fully bitwise reproducible planetary histories that result from this modified version of {\sc rebound} significantly simplify the problem of simulating planetary migration. 
As noted above in Section~\ref{ss:scheme}, it can be challenging to find initial conditions for the planets that result in post-migration orbits similar to their present ones; this is true in both full N-body migration simulations and simplified parameterized migration simulations.
We find that, especially for slow migration, the planets often cross mutual mean motion resonances slowly enough for eccentricities to be significantly excited; this can sometimes lead to close-encounters between Uranus and Saturn or Uranus and Neptune, which further excite the system. 
Because the exact excitation that results from resonances depends on the phases of the planets when they encounter them, even tiny changes in the initial conditions can lead to wildly different outcomes. 
With integrators such as SWIFT, we have even found that running identical initial conditions on machines with different processors can produce different results, even if the same compiler is used. 
This presents a challenge when running simulations on computer clusters with multiple node architectures; it also means that numerical experiments cannot be tested with additional test particles or for extended integration lengths if the original machine is no longer available. 
The ability to remove machine- and compiler-dependent round off error as a source of chaos in the numerical integrator \citep[as done by][]{Rein:2017} is particularly helpful for planet migration simulations.

We investigated four migration e-folding timescales: $\tau_a~=$ 5 Myr, 10 Myr, 30 Myr, and 50 Myr. 
For each timescale, we ran a test suite of $\sim~500$ simulations of just the four migrating giant planets. 
Jupiter and Saturn were initialized on orbits with their current eccentricities and inclinations but semimajor axes of 5.4~au and 8.8~au, respectively. Uranus and Neptune were initialized on orbits of semimajor axes randomly chosen from the ranges $16.4\pm0.035$~au and $24\pm0.035$~au, respectively; their inclinations were set to their current values, and their eccentricities were set randomly in the range $e=0.005\pm0.002$. 
We integrated each set of initial conditions to $t~=~700$~Myr and calculated the average semimajor axis value and the eccentricity ranges of all 4 planets over the last 100 Myr of the simulation. 
For each migration timescale, we then selected a set of initial conditions that resulted in a planetary system that best matched the current solar system in terms of semimajor axis ratios and eccentricity ranges. 
The final eccentricity and inclination ranges of the planets in these simulations (taken over the last 100~Myr of the simulations) are shown in the top row of panels of Figure~\ref{f:planets}.

For each combination of migration timescale and planets' initial conditions, we then re-integrated the planets along with $3.5\times10^5$ massless test particles. The test particles were given initial semimajor axes in the range $24.5-30$~au (from just outside Neptune's initial orbit to just inside its current orbit); 
their eccentricities and inclinations were randomly drawn from a Raleigh distribution of width 0.025 (where the units are in radians for the inclinations). 
These test particle initial conditions are very similar to those used in \citet{Nesvorny:2015}.
The planets and test particles were integrated to $t=700$~Myr. 
This 700 Myr integration length was chosen as a compromise between conserving computational time and reaching a post-migration final orbital distribution of test particles that are stable on $\sim$~gigayear timescales for comparison with the present-day observed Kuiper belt.
(Our simulations required $\sim2.5\times10^4$ cpu hours per migration timescale; a total of $\sim10^5$ cpu hours  were used in this study on the Ocelote cluster maintained by UA Research Computing High Performance Computing at the University of Arizona.) 
It is pertinent to note that $\lesssim15\%$ of the Plutino population is expected to have leaked out of Neptune's 3:2 resonance in the past $\sim3.5$~gigayears, and that this loss is not strongly inclination dependent~\citep{Nesvorny:2000,Tiscareno:2009}.
Therefore, our simulation length is sufficient to test how the planet migration rate affects the final Plutino inclination distribution, the dynamical class of KBOs for which \citealt{Nesvorny:2015} found the strongest dependence on migration rate. 
The 700~Myr timescale is also sufficient to investigate the inclination distribution of the hot classical population because the inclination distribution of this population does not evolve significantly on long timescales \citep[e.g.,][]{Volk:2011}. 
The only region of the classical belt that has strongly inclination-dependent stability is the inner region near the $\nu_{18}$ secular resonance \citep[e.g.,][]{Kuchner:2002}; in this region, the instability timescale is much shorter than 700~Myr.

As discussed later in Section~\ref{s:discussion}, our simulations suggest that the strengths of the giant planets' inclination secular modes during and after migration have a significant influence on the inclination distribution of the post-migration Kuiper belt. 
When selecting the planets' initial conditions for the set of simulations described above, we had focused on matching the final eccentricities of the planets rather than on matching the inclinations.
At the end of these four simulations described above, Neptune's inclination is in the range 1.3$^\circ$--2.0$^\circ$, slightly larger than in the real solar system (top row of panels in Figure~\ref{f:planets}). This results in more power being associated with Neptune's dominant inclination secular frequency than in the real solar system.
This realization led us to perform an additional set of migration simulations for each migration timescale. 
The initial conditions for the planets in this new set of simulations were chosen the same way as described above, except that we reduced the initial mutual inclinations of the giant planets to try to reduce Neptune's final inclination to more closely match the real solar system.
After running a suite of $\sim500$ planets-only migration simulations, we selected initial conditions that resulted in a better match for Neptune's final inclination; this also resulted in slightly smaller inclinations for other planets compared to the real solar system.
The final eccentricity and inclination ranges of the planets in these simulations (taken over the last 100~Myr of the simulations) are shown in the bottom row of panels in Figure~\ref{f:planets}.
These planetary initial conditions were then re-run with $1.5-2.5\times10^5$ massless test particles with initial conditions as described above.  
The second set of simulations are labeled `B', while the first set of simulations are labeled `A'.

\subsection{Simulation Results}\label{s:results}

At the end of each migration simulation, we identified the test particles representative of the Plutino population by comparing test particles' time averaged semimajor axes for the last 50~Myr of the simulation to the expected semimajor axis for the resonance. 
We also identified the population of test particles in the "hot classical" region, defined as those with semi-major axes $40<a<48$~au and perihelion distances above 35~au and not obviously in resonance with Neptune \citep[consistent with][]{Nesvorny:2015}; we discarded test particles with average semimajor axes indicative of libration within the 8:5, 5:3, 7:4, and 2:1 mean motion resonances from this hot classical population.
The capture efficiencies and median inclinations for the resulting Plutino and hot classical populations from each simulation are given in Table~\ref{t:sims}.

\begin{deluxetable*}{cc|clcl}
\tablecaption{Summary of simulation results\label{t:sims}}
\tablehead{
$\tau_a$ & simulation & median $i_{HC}$ & $f_{HC}$ & median $i_{3:2}$ & $f_{3:2}$
}
\startdata
50 Myr & A & 26$^\circ$ & $\sim5\times10^{-4}$ & 13$^\circ$ &  $\sim1\times10^{-3}$ \\ 
50 Myr & B & 17$^\circ$ & $\sim1\times10^{-4}$ & 10$^\circ$ &  $\sim1.5\times10^{-3}$ \\ 
30 Myr & A & 27$^\circ$ & $\sim6\times10^{-4}$ & 13$^\circ$ &  $\sim6\times10^{-4}$ \\ 
30 Myr & B & 18$^\circ$ & $\sim2.5\times10^{-4}$ & 9$^\circ$ &  $\sim1.5\times10^{-3}$ \\ 
10 Myr & A & 20$^\circ$ & $\sim8\times10^{-4}$ & 18$^\circ$ &  $\sim5\times10^{-4}$  \\
10 Myr & B & 13$^\circ$ & $\sim5.5\times10^{-4}$ & 8$^\circ$ &  $\sim1\times10^{-3}$  \\
5 Myr  & A & 21$^\circ$ & $\sim7\times10^{-4}$ & 10$^\circ$ &  $\sim2\times10^{-4}$ \\ 
5 Myr  & B & 18$^\circ$ & $\sim3\times10^{-4}$ & 7$^\circ$ &  $\sim2.5\times10^{-4}$ \\ 
\enddata
\tablecomments{
The median inclination and fraction of initial test particles in the 3:2 and hot classical populations ($f_{3:2}$ and $f_{HC}$) at the end of the 700 Myr simulations; all inclinations are measured relative to the plane defined by the planets' total angular momentum.
As described in Section~\ref{s:sims}, the simulations labeled `A' resulted in Neptune having an inclination slightly larger than in the current solar system; the planets inclinations in the simulations labeled `B' are slightly lower than in the current solar system (see also Figure~\ref{f:planets}).
}
\end{deluxetable*}

\subsubsection{Capture efficiencies for the hot classical and Plutino populations}

We find rather small capture efficiencies for the hot classical region, similar to those found by \citet{Nesvorny:2015}. 
For example, with a $\tau_a=30$~Myr migration timescale and a similar initial Neptune orbit, \citet{Nesvorny:2015} found a hot classical capture efficiency of $\sim~2~\times~10^{-4}$ compared to our efficiencies of $\sim6\times~10^{-4}$ and $\sim~2.5~\times~10^{-4}$ for our two $\tau_a=30$~Myr simulations. 
Note, however, that the \citet{Nesvorny:2015} capture efficiencies reflect 4~Gyr of evolution rather than our 700~Myr.
The results of our two sets of simulations (`A' and `B' in Tables~\ref{t:sims}) show that the capture efficiency in the hot classical region varies significantly for the same migration timescale with different initial conditions. 
For fixed migration timescales, the efficiencies vary by a factor of 1.5-5, while across all simulations (and migration timescales) the efficiency only varies by a factor of 8.
This shows that capture efficiencies depend on more than just the migration timescale and that the timescale dependence is weak, or at least not so strong as to overwhelm these other dependencies.

The Plutino capture efficiencies in our simulation are also broadly similar to those of \citet{Nesvorny:2015}, although ours are slightly lower when accounting for the shorter total integration time. 
For $\tau_a=10$~Myr and $\tau_a=30$~Myr, \citet{Nesvorny:2015} described capture efficiencies in the range $\sim0.5-1\times10^{-3}$ compared to efficiencies in the range $\sim0.5-1.5\times10^{-3}$ in our similar simulations.
While \citet{Nesvorny:2015} found that slower migration timescales decreased the efficiency of capture into the 3:2 resonance (for simulations with Neptune starting near $\sim24$~au, like in our simulations), we find a weak trend in the opposite direction; for both our `A' and `B' simulations, the $\tau_a=5$~Myr simulations resulted in capture efficiencies ~2-5 times lower than for the longer migration timescales.

We note that the capture efficiencies into the Plutino and hot classical regions depend on many factors, including some not explicitly investigated here. 
Resonant capture efficiencies depend not only on Neptune's migration speed but also the eccentricities of the test particles \citep[e.g.,][]{Hahn:2005} and the smoothness of migration \citep[e.g.,][]{Nesvorny:2016,Kaib:2016}; the smoothness of migration can also affect the capture efficiency into the hot classical population.
For the hot classical population, \citet{Dawson:2012} found that Neptune's apsidal precession rate also affects the probability of capturing scattered test particles onto stable orbits.
All of these factors mean that the capture efficiency can vary widely even for a single planetary migration timescale, as seen in Table~\ref{t:sims}. 
The relative capture efficiency between the Plutinos and hot classical population also vary widely; our two different $\tau_a=30$~Myr simulations produce ratios of Plutinos to hot classical KBOs of 6 and 1.
Thus outcomes from single sets of initial conditions, such as capture efficiency, population ratios, and orbital parameters, can be difficult to generalize.

Our simulations do confirm that a scattering origin for the resonant populations is much less efficient than sweep up from a dynamically cold source considered in classical resonance sweeping scenarios \citep[e.g.,][]{Malhotra:1995,Hahn:2005}.
We list our capture efficiencies in Table~\ref{t:sims} as both a check for consistency with similar migration simulations in the literature and to emphasize the large number of test particles  needed in simulations to capture even moderate numbers in the hot classical and Plutino populations.

\subsubsection{Inclination distributions in the hot classical and Plutino populations}

\begin{figure*}
\centering
   \begin{tabular}{c c}
   \includegraphics[width=3.25in]{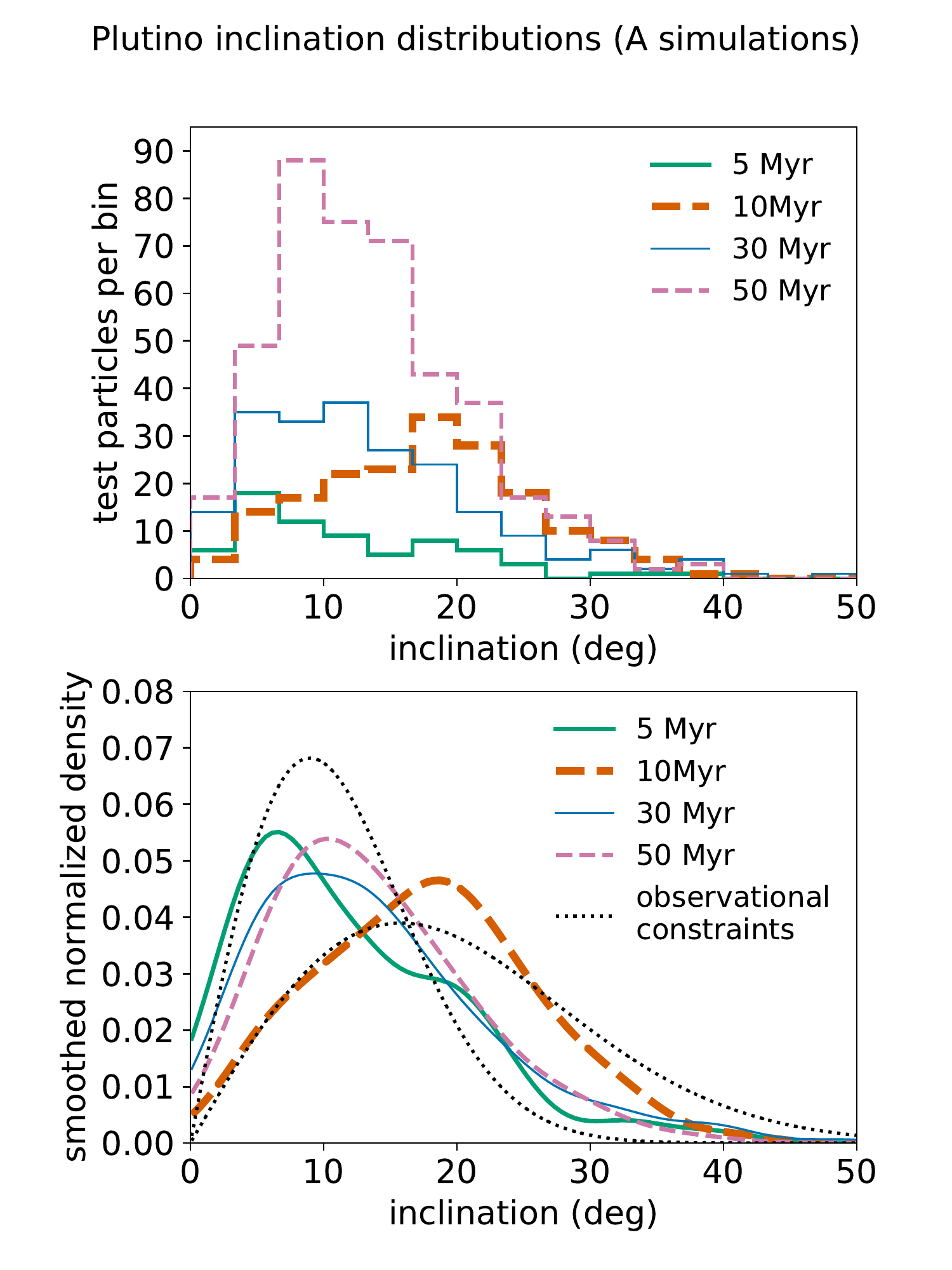} &
   \includegraphics[width=3.25in]{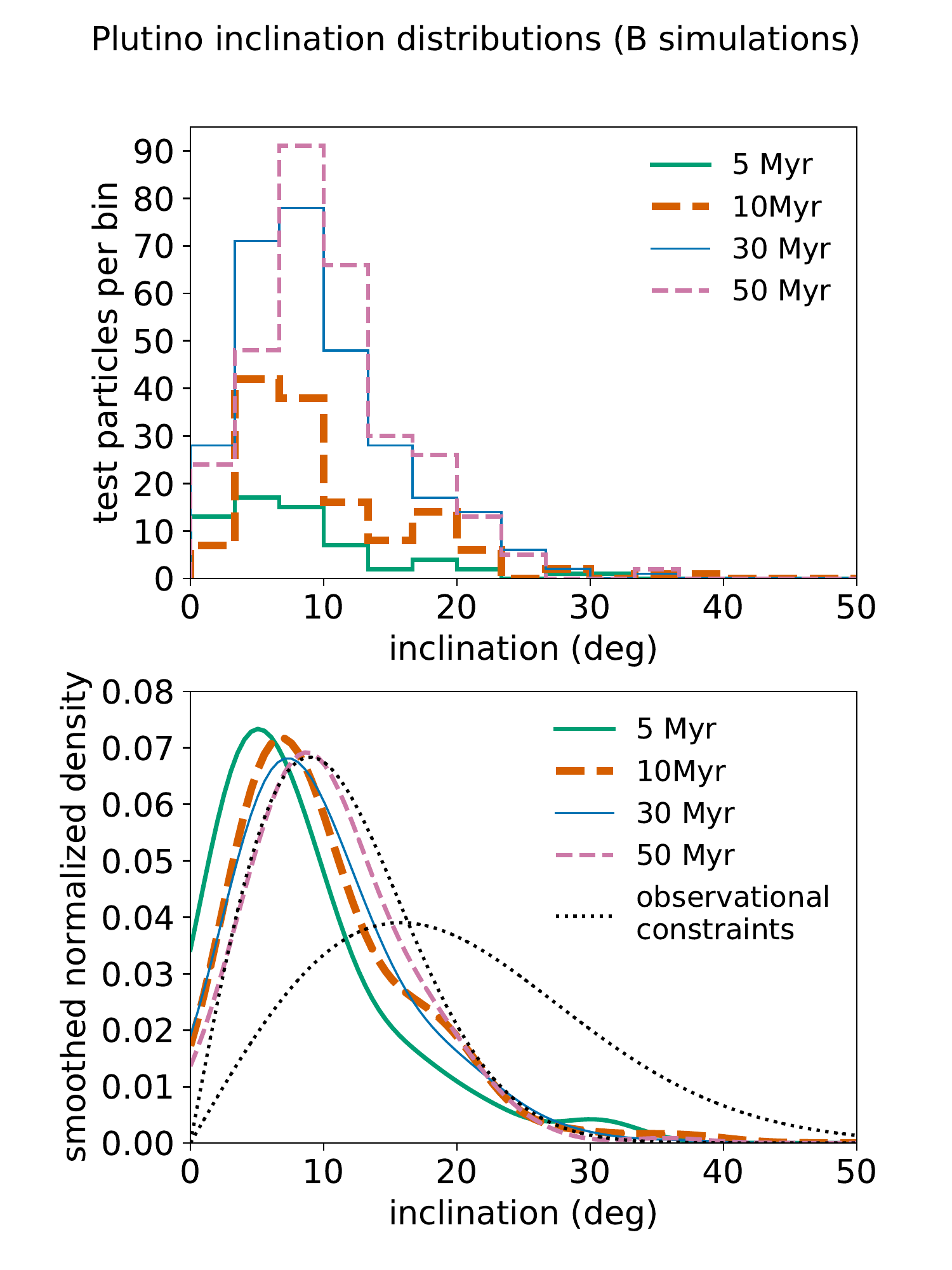} \\
   \end{tabular}
   \caption{
   Top panels: Histograms of the final inclinations (at $t=700$~Myr) of test particles captured in Neptune's 3:2 resonance for our four different migration timescales in simulation set `A' (top left) and simulation set `B' (top right). 
   Bottom panels: normalized and smoothed kernel density estimates for the same inclination distributions;  also plotted with black dashed lines are the 
   95\% confidence limits on the Plutino population's intrinsic inclination distribution, assuming it has the form $f(i)\propto \sin(i)\exp{\frac{-i^2}{2 \sigma_i^2}}$\citep{Volk:2016}. 
   These smoothed distributions are shown to better illustrate the differences between the simulations. \label{f:plutino-incs} } 
\end{figure*}

Figure~\ref{f:plutino-incs} shows the inclination distributions of the final Plutino populations for our four migration timescales; the distributions from the `A' simulations (in which Neptune's inclination is slightly larger than in the real solar system) are shown in the left panels, and the distributions from the `B' simulations (in which the planets have slightly smaller inclinations) are shown in the right panels.
For each simulation, we have measured the inclinations of the test particles relative to the plane defined by the total angular momentum of the planets. 
As noted above, the resonance capture efficiency is quite low, resulting in a somewhat noisy inclination distribution for each individual simulation despite the large initial number of test particles.
In the $\tau_a=5$~Myr simulations, at the end ($t=700$~Myr) we had only $\sim70$ Plutinos in both the `A' and `B' simulations; each of the longer migration timescale simulations ended with $\sim140-500$ Plutinos.
While the inclination distributions are slightly noisy, our results are sufficient to observe a counter example to \citet{Nesvorny:2015}'s result that the peak in the Plutino inclination distribution increased markedly with increasing migration timescale, shifting from $\sim10^\circ$ for $\tau_a=10$~Myr to $\sim20^\circ$ for $\tau_a=30$~Myr (see his Figure 9).
We find no such trend in our simulations. Our `A' and `B' simulations with migration timescale $\tau_a=10$~Myr yield both the highest median Plutino inclination and a near-tie for lowest median inclination. 
Our `B' simulations have a slight trend of increasing median Plutino inclinations with increasing migration timescale, but it is very weak with median inclinations only increasing from $7^{\circ}$ to $10^{\circ}$ over an order of magnitude in $\tau_a$; additionally, the difference between the inclination distributions for $\tau_a=10$~Myr and $\tau_a=30$~Myr `B' simulations is not statistically significant.
Overall our simulations show that there is as much variation in the particle inclination distributions when comparing multiple runs with the same migration timescale but slightly different initial conditions of the planets as when comparing runs with different migration timescales.

In the bottom panels of Figure~\ref{f:plutino-incs}, we plot smoothed inclination distributions using a normalized kernel density estimate for each simulated set of Plutinos. 
(Note that any physical inclination distribution should go to zero at $i=0$, which is not the case for these smoothed distributions; these smoothed distributions are shown primarily to more easily compare the simulations.) 
We also plot observationally derived estimates of the intrinsic inclination distribution of the Plutinos (dotted black curves in Figure~\ref{f:plutino-incs}) for comparison to our simulation results; these are from \citet{Volk:2016} who modeled the observed set of Plutinos discovered in the first quarter of the Outer Solar System Origins Survey \citep{Bannister:2016} and determined that their intrinsic inclination distribution could be acceptably modeled by a function, $f(i) \propto \sin(i) \exp{(i^2/2\sigma_i^2)}$, with parameter $\sigma_i \approx 9-13^\circ$ (consistent with previous estimates, e.g., \citealt{Gladman:2012,Gulbis:2010}). 
We see that the results of our `A' simulations are generally similar to the estimated intrinsic inclination distributions while our `B' simulations result in Plutinos with typical inclinations slightly smaller that the observations suggest.

\begin{figure*}
\centering
   \begin{tabular}{c c}
   \includegraphics[width=3.25in]{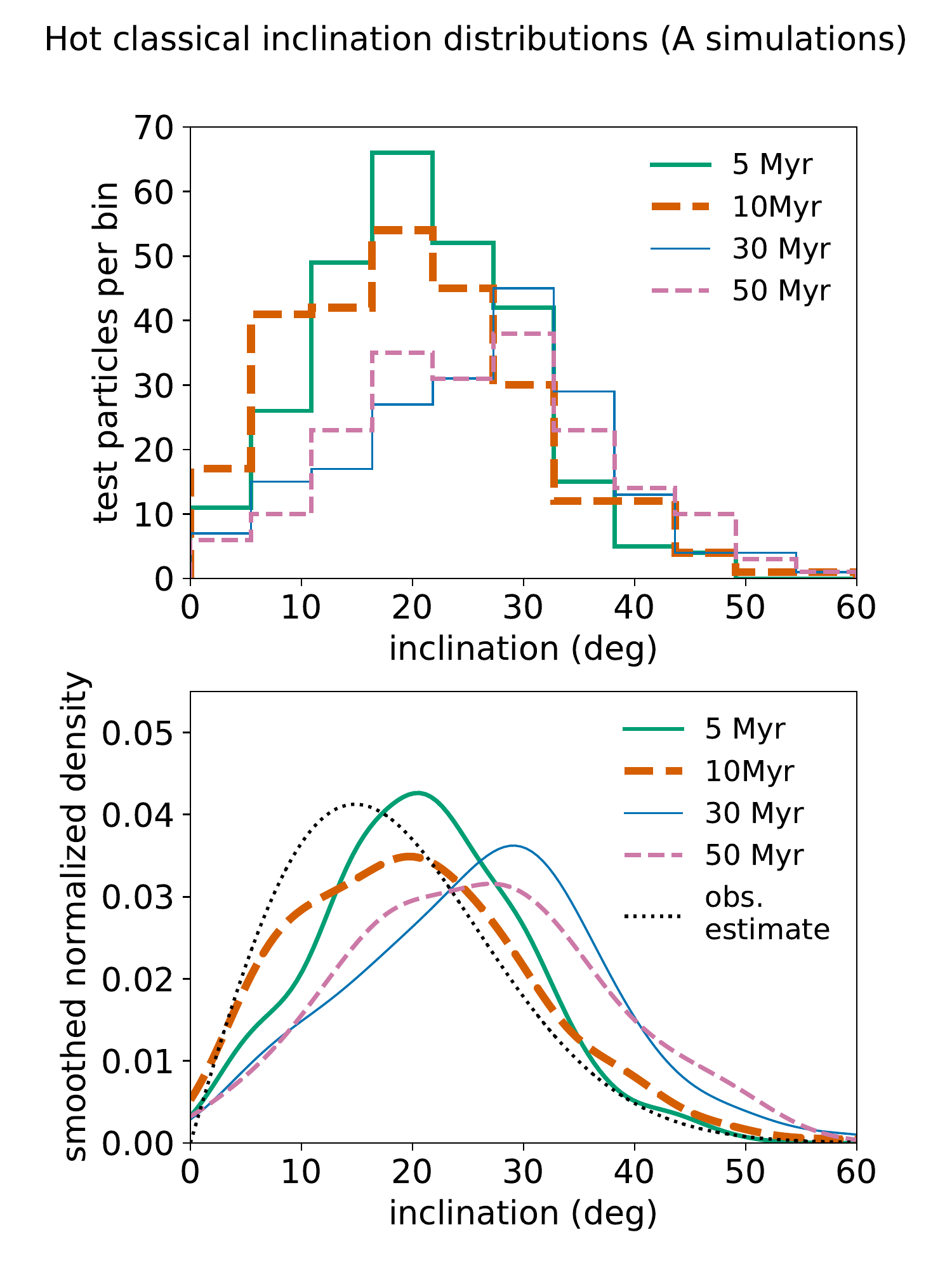} &
   \includegraphics[width=3.25in]{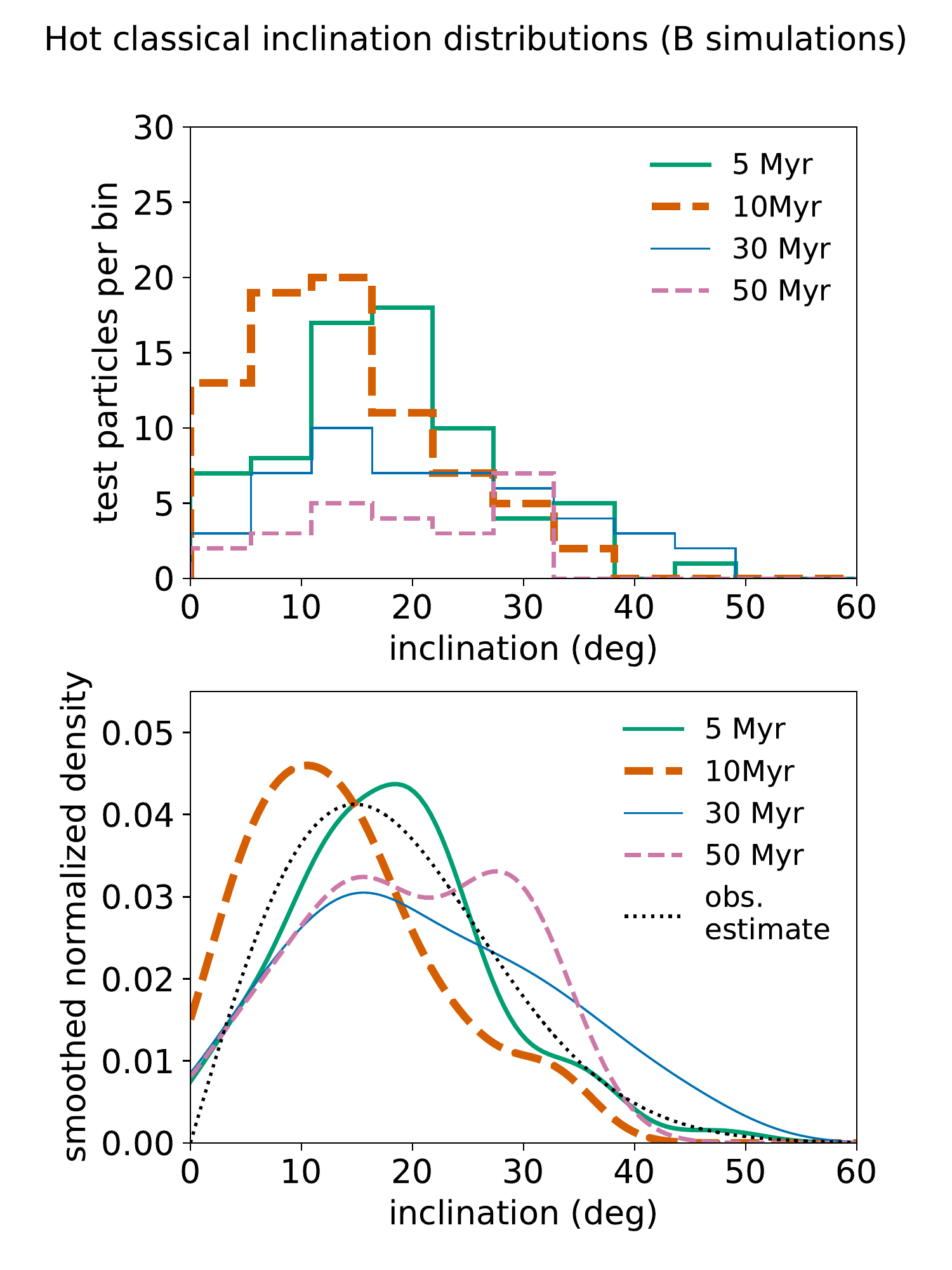} \\
   \end{tabular}
   \caption{Top panels: Histograms of the final inclinations (at $t=700$~Myr) of test particles in the hot classical region for our four different migration timescales in simulation set `A' (top left) and simulation set `B' (top right); note that the top panels have different y-axis ranges. Bottom panels: Normalized and smoothed kernel density estimates for the same inclination distributions;  also plotted with black dashed lines are observationally derived best-fit functions ($\sin i$ times a Gaussian) for the dynamically hot population \citep{Petit:2017}. \label{f:hc-incs}}
\end{figure*}

Figure~\ref{f:hc-incs} shows the inclination distribution for the test particles in the hot classical region at the end of the simulations for each simulated migration timescale.
Our hot classical test particles have inclination distributions that are slightly broader and extend to higher inclinations than our Plutino distributions.
We again note that there is not a clear trend between peak inclination and migration timescale. 
The inclination distribution for our $\tau_a=30$~Myr  migration timescale simulation `A' peaks in the range $\sim20-30^\circ$, which is consistent with \citet{Nesvorny:2015}'s result (his Figure~9); our $\tau_a=30$~Myr migration timescale simulation `B' peaks at slightly lower inclination, but is of similar width. 
For comparison, we show \cite{Petit:2017}'s estimate for the intrinsic inclination distribution of the hot population in the bottom panels of Figure~\ref{f:hc-incs}.
All of our simulations produce fairly broad inclination distributions of the hot classical population, similar to the estimated intrinsic distribution.

We compared our simulated inclination distributions to each other using the Anderson-Darling test\footnote{We use the version of this test described by NIST:  \url{https://www.itl.nist.gov/div898/handbook/eda/section3/eda35e.htm}, also described in Appendix~1 of \citet{Volk:2016}.}.
For the Plutinos shown in Figure~\ref{f:plutino-incs}, most of the distributions are statistically distinguishable at 95\% confidence except when comparing the $\tau_a=5$~Myr `A' simulation to either the `A' or `B' $\tau_a=30$~Myr simulations and when comparing the $\tau_a=10$~Myr `B' and $\tau_a=30$~Myr `B' simulations.
For the hot classical inclination distributions shown in Figure~\ref{f:hc-incs}, the `A' simulations are distinct from each other with the exception of the $\tau_a=30$~Myr and $\tau_a=50$~Myr distributions.
There are fewer captured hot classical test particles in the `B' simulations (due partly to lower capture efficiencies and partly due to smaller numbers of initial test particles), so these distributions are not as distinct. 
The $\tau_a=10$~Myr `B' simulation is statistically distinguishable from the other `B' simulations, but the other timescales are not distinguishable from each other at 95\% confidence. 
The $\tau_a=5$~Myr `B' and $\tau_a=10$~Myr `B' hot classical inclination distributions are statistically different from all of the `A' hot classical distributions; the $\tau_a=30$~Myr `B' and $\tau_a=50$~Myr `B' inclination distributions are not distinguishable from the `A' simulations because few test particles were captured into the hot classical region.
 Considering that most of our simulations produced statistically distinct inclination distributions, particularly in the final Plutino populations, we should have easily identified the strong trend seen in \citet{Nesvorny:2015}'s simulations if that trend were present in our simulations.

\section{Discussion}\label{s:discussion}

Our simulation results are in some conflict with \citet{Nesvorny:2015}'s.  
To identify the cause(s) of the discrepancies, we first note some important differences between the simulated migration scenarios.
\citet{Nesvorny:2015} forced only Neptune to migrate in the simulations (keeping the orbits of the other giant planets fixed), while we have all four giant planets migrating. 
Another major difference is that \citet{Nesvorny:2015} employed an eccentricity and inclination damping scheme for Neptune (in addition to the torque to migrate its semi-major axis), while we include only the torque required to migrate the planets in semimajor axis. 
These two differences lead to some differences in the secular architecture of the simulated planetary system, both during and after migration, which are likely to be important for shaping the inclination distributions of the hot classical and Plutino populations.

In the current solar system, the evolution of the four giant planets' orbital planes (i.e., their orbital inclinations, $i$, and longitudes of ascending node, $\Omega$) can be described fairly accurately using linear secular theory \citep[see, e.g.,][]{Murray:1999SSD}. 
Briefly, the premise of linear secular theory is that the planets' mutual gravitational perturbations can be modeled as though the mass of each planet were spread out in a ring along its eccentric, inclined orbit; this amounts to assuming that the perturbations between the planets can be averaged over orbital timescales.
The rings representing each planet perturb each other, causing quasi-periodic variation of the planets' eccentricity vectors and inclination vectors while the planets' semimajor axes remain fixed.
To linear order, the evolution of the planets' orbital planes and their orbital eccentricities are independent.  Since we are investigating the inclination distributions of the Plutino and hot classical populations, we focus here on the linear secular theory of the planets' inclinations.
Each planet's orbital plane is described by an inclination vector ($\sin i \sin \Omega$, $\sin i \cos \Omega$). 
The equations of motion for the linear secular perturbations of the planets allow a solution  in which the time evolution of the planets' inclination vectors can be described by a sum over the system's inclination eigenvectors and eigenfrequencies.
The solution for our four giant planets yields three non-zero eigen-modes called the secular inclination modes.
In the notation of \citet{Murray:1999SSD}, these are the $f_6$, $f_7$, and $f_8$ modes; in the current solar system they correspond to nodal regression periods of $\sim0.05$, $\sim0.45$, and $\sim1.9$ Myr, respectively.
The frequency of each of these modes (to lowest order) depends only on the masses and semimajor axes of the planets.
Because our simulations are designed to have the planets end very near their current semimajor axes, the mode frequencies at the end of each simulation are nearly identical and match the mode frequencies of the current solar system quite well.
For each giant planet, the  relative contribution of these three modes to the time-evolution of its inclination vector (i.e., the mode amplitudes calculated from the system's eigenvectors) is determined by initial conditions.
We can calculate the mode amplitudes in the current solar system using the observed planetary inclinations.
The combination of mode frequencies and mode amplitudes is what we refer to as the secular architecture of the giant planets.

Linear secular theory can also be used to describe how the inclinations of massless test particles are affected by the planets. 
 A test particle's inclination vector can be described as a sum of two components: a free inclination and a forced inclination, determined by initial conditions and the secular architecture of the giant planets.
The amplitude of the forced inclination depends on the precession rate of the test particle's free inclination and the planets' secular modes and mode amplitudes. 
The free precession rate is determined by the averaged perturbations from the giant planets and depends only on the planets' masses and semimajor axes combined with the test particle's semimajor axis.
The amplitude of the test particle's forced inclination is determined by a weighted sum over each of the planetary  secular modes; the contribution from each mode is proportional to the mode amplitudes (calculated form the perturbations amongst the planets) and inversely proportional to the difference between the test particle's free precession rate and the mode frequencies.
Very large forced inclinations occur where the free precession  frequency matches one of the secular mode frequencies.
The locations in test particle semimajor axes where this occurs are referred to as secular resonances (see \citealt{Knezevic:1991} for a map of secular resonances in the solar system).

In the current solar system, the free precession periods of test particles in the outer solar system range from $\sim0.25$~Myr at 20~au (just outside Uranus's orbit) to $\sim5$~Myr at 50~au (the outer edge of the classical Kuiper belt).
Thus, only the $f_7$ and $f_8$ contribute significantly to the forced inclination of a test particle in the Kuiper belt region.
The influence of the $f_8$ mode is particularly apparent in the current Kuiper belt; the secular resonance associated with the $f_8$ mode, referred to as the $\nu_{18}$ secular resonance, is located in the $a=40-42$~au range \citep[see, e.g.,][]{Chiang:2008}, helping to define the inner edge of the classical Kuiper belt.

We can use a linear secular analysis of our migration simulations to help understand how the secular architecture of our simulated planetary systems might be affecting the test particle inclinations in our simulations. 
We can use the theory to calculate the modes and mode amplitudes of the planets in our simulations to see how they vary between simulations and how they compare to the current solar system.
However, there are several caveats that we need to acknowledge before discussing this analysis.
First is that a linear secular description of the planets' interactions does not include the effects of the extra migration forces in the simulation.
At each time point being considered, we are effectively treating the system as having "frozen" semi-major axes at each point in time.
In our slower migration cases, the planets' orbits are changing in semimajor axis on timescales longer than the outer planets' secular timescales which are $\lesssim2$~Myr; for the shorter migration timescales, the evolution toward the end of migration is slow enough to be much slower than these secular timescales, but this separation of migration timescale and secular timescale is not so large at the beginning of the simulations. 
We also recognize that linear secular theory is not a good approximation of the planets' evolution if they approach and/or cross mutual mean motion resonances as they migrate.
In such cases, the assumption that we can average over orbital timescales breaks down.

\begin{figure}[htbp]
\centering
   \includegraphics[width=3in]{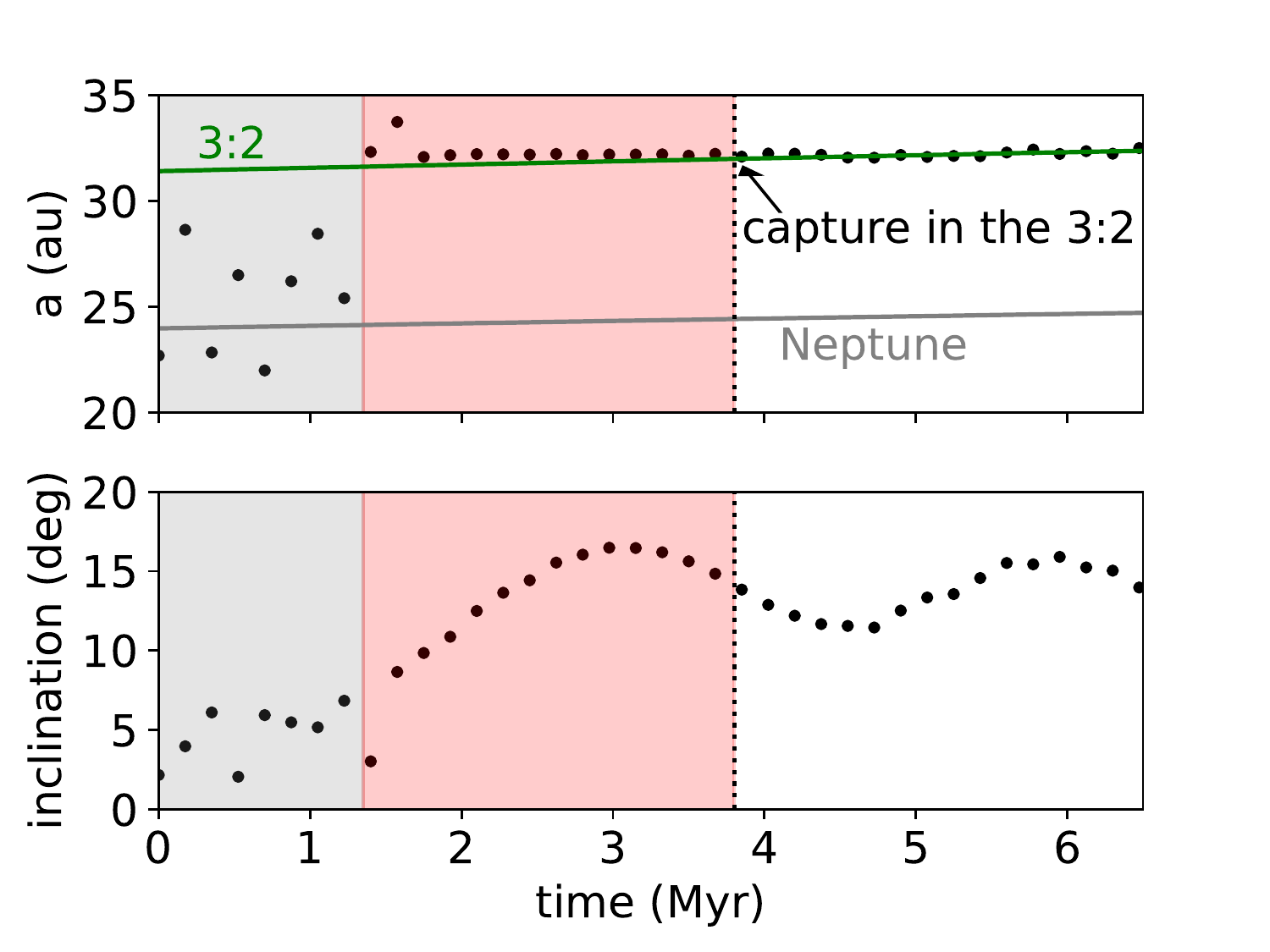}
   \caption{Example of secular inclination excitation for a test particle in the $\tau_a=50$~Myr `A' simulation. The top panel shows the semimajor axis of the test particle (black dots) compared to the location of Neptune (gray line) and the 3:2 resonance (green line); the bottom panel shows the test particle's inclination. The gray shaded region indicates the time period where the test particle's evolution is dominated by scattering events with Neptune. The red shaded region indicates the time period where a marked secular increase in the test particle's inclination occurs. The vertical dashed line indicates the time where the particle is captured in the 3:2 resonance. \label{f:secular-inclination-excitation}} 
\end{figure}

The evolution of test particles in the simulations are not expected to be dominated by secular perturbations, especially at the beginning of migration. 
The orbit averaging assumption is not valid when test particles are on planet-crossing orbits.
Because all of the test particles must be scattered outward to end up in the Plutino or hot classical regions, they are clearly subject to perturbations that are not secular in nature.
However, their orbital inclinations between or after scattering events will still be influenced by secular forcing.
In particular, the semimajor axis of the 3:2 resonance is close to the semimajor axis of the $\nu_{18}$ inclination secular resonance where significant secular forcing can occur on timescales comparable to the mode period, so the inclinations of the Plutino population could be particularly sensitive to variations in the secular architecture of the planets.
For example, Figure~\ref{f:secular-inclination-excitation} shows the inclination and semimajor axis evolution for a test particle at the beginning of the $\tau_a=50$~Myr `A' simulation. 
In the first million years, the particle experiences a few degrees of inclination excitation as it scatters with Neptune. Then the particle is scattered to a semimajor axis just beyond the 3:2 resonance, which is near the estimated location of the $\nu_{18}$; this results in the test particle's inclination increasing by $\sim10^\circ$ over a $\sim1$~Myr period. 
Neptune's continued migration then results in the 3:2 resonance catching up with the test particle and capturing it into resonance, locking it into its higher inclination orbit.
Thus, despite the limitations, a linear secular analysis of the simulations can yield useful insights into the resulting test particle distributions and illustrate the potential for secular forcing to be an important source of inclination excitation.

\begin{figure*}[htbp]
\centering
   \begin{tabular}{c c}
   \includegraphics[width=3in]{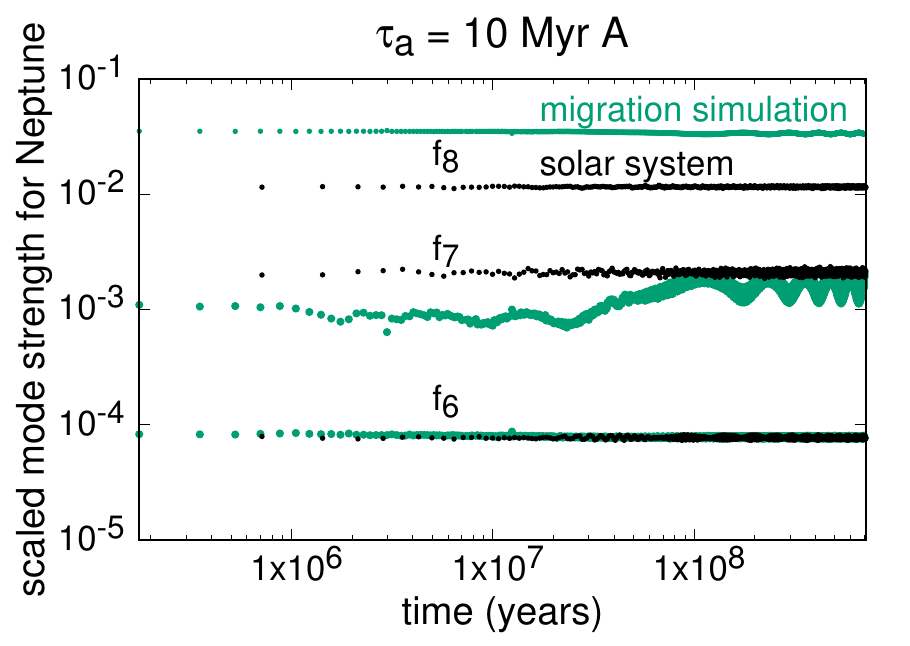} &
   \includegraphics[width=3in]{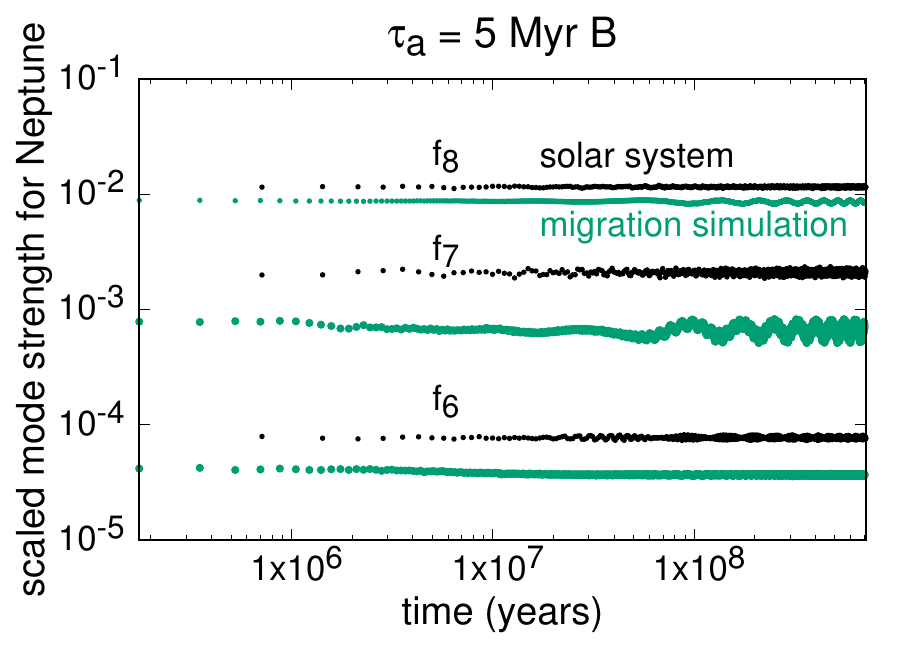} \\
   \end{tabular}   
   \caption{   Estimated amplitudes for the three inclination secular modes in Neptune's orbital inclination for the $\tau_a=10$~Myr `A' simulation (green, left panel) and the $\tau_a=5$~Myr `B' simulation (green, right panel) compared to the same mode strengths for the current solar system (black).   \label{f:mode-power}  }
\end{figure*}

We use linear secular theory to estimate the amplitudes of the three secular modes in our simulations and to compare them to the mode amplitudes in the real solar system.
We do this by taking snapshots of our simulated systems at various points. 
These points in time define the semimajor axes of the planets (the masses of the planets are fixed to their observed values throughout the simulations) from which the mode frequencies can be calculated. 
Then, just as the observed inclinations of the planets are used to calculate the mode amplitudes in the real solar system, we use the instantaneous inclination vectors of the planets at these times in the simulation to determine the appropriate mode amplitudes.
For comparison, we performed the same calculations for a 700~Myr simulation of the current solar system's giant planets. 
The black lines in Figure~\ref{f:mode-power} show the calculated amplitudes of the $f_6$, $f_7$, and $f_8$ in Neptune's inclination evolution in this simulation of the current solar system.
The fact that the calculated mode amplitudes are relatively flat shows that linear secular theory provides a good description of Neptune's inclination evolution; if Neptune's simulated inclination vector had additional frequencies with significant amplitudes besides the three predicted by the simplified theory, the  mode amplitudes recalculated at each time-point in the simulation would not match.
Figure~\ref{f:mode-power} shows, in green, the same mode amplitudes for Neptune from two of our migration simulations.
 The larger variations of these mode amplitudes in the migration simulations compared to the simulation of the non-migrating current solar system reflects the fact that linear secular theory (as described above) does not account for mean motion resonances that can produce significant perturbations as the planets migrate.
Nevertheless, for each simulation, the dominant mode in Neptune's inclination evolution (the $f_8$ mode) is relatively stable, especially at the end of the simulations when migration has finished.

As Figure~\ref{f:mode-power} demonstrates, our simulations  lead to a final secular architecture of the giant planets that is not a perfect simile of the real solar system.
In our `A' simulations, the amplitude of $f_8$ mode is typically larger than in the current solar system, even when the power associated with the other modes is fairly well matched (as in the left panel of Figure~\ref{f:mode-power}). 
Because the $f_8$ is the primary mode associated with Neptune's inclination, this results in the simulated Neptune having an inclination slightly too large compared to the real solar system (see Figure~\ref{f:planets}).
The power associated with the $f_8$ mode in our `B' simulations better matches that of the real solar system (corresponding to a better match with Neptune's real inclination), although the other mode amplitudes are typically slightly smaller (as in the right panel of Figure~\ref{f:mode-power}).

\begin{figure*}[htbp]
\centering
   \begin{tabular}{c c}
   \includegraphics[width=3in]{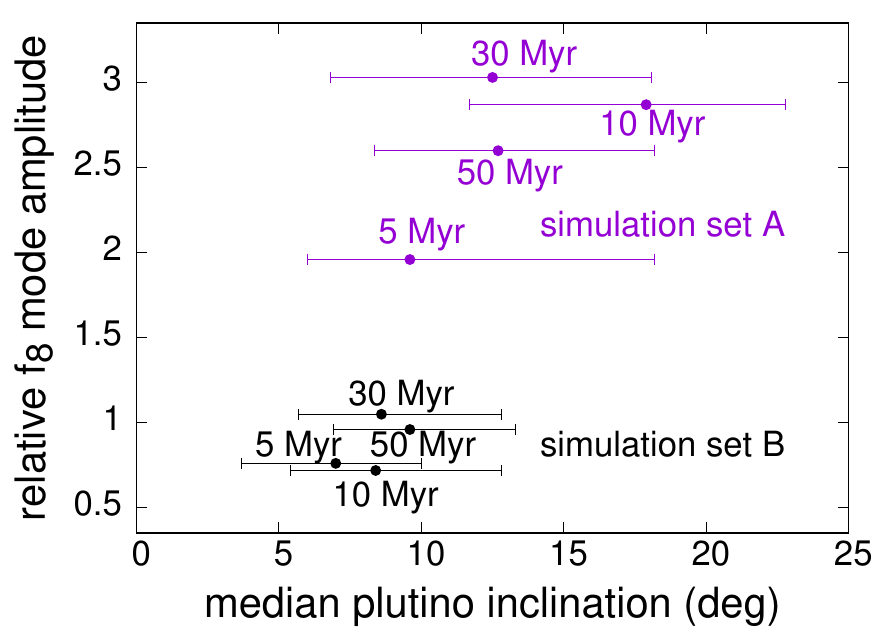} &
   \includegraphics[width=3in]{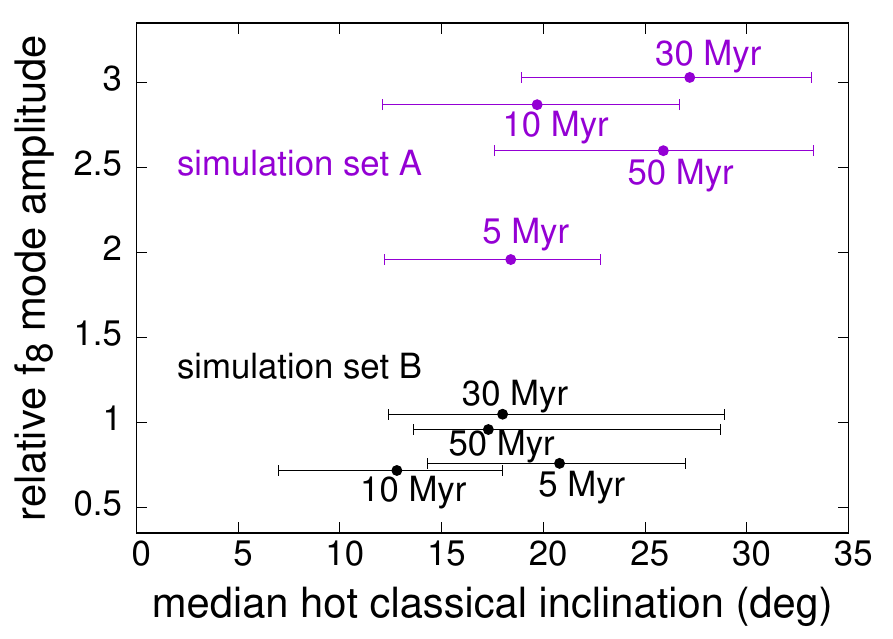} \\
   \end{tabular}   
   \caption{   The ratio of the estimated amplitude of the $f_8$ inclination mode in Neptune's orbit at the end of the simulations to its amplitude in the current solar system vs. median inclination in the final Plutino (left panel) and hot classical populations (right panel) in the simulations.
   The horizontal bars represent the 25-75\% inclination range in each population's cumulative inclination distribution.
    The `A' simulations are shown in purple and the `B' simulations in black (labeled by $\tau_a$).\label{f:mode-ratios}   }
\end{figure*}

This very simple analysis of the $f_8$ mode amplitudes in our simulations hints at a plausible explanation for the trends in the inclination distributions in Figure~\ref{f:plutino-incs}. 
For each simulation, we calculate the average amplitude of the $f_8$ mode in Neptune's orbit at the end of the simulation and ratio it to the same amplitude for the current solar system.
We plot these mode amplitude ratios and the inclinations of the Plutino and hot classical populations
in Figure~\ref{f:mode-ratios}, observing the following points. First, the `A' simulations have mode amplitudes that are  $\sim2-4$ times larger than the `B' simulations. Second, the `A' simulation Plutino populations (left panel of Figure~\ref{f:mode-ratios}) also have higher inclinations than the Plutinos in the `B' simulations.  The inclination widths, defined as the middle 50\% of the inclination distribution, in the `A' simulation Plutinos are $\sim2$ times larger than for the `B' simulation Plutinos; the median inclinations in the `A' simulations are also up to $\sim3$ times larger than in the `B' simulations.

These trends are much weaker in the hot classical population (right panel of Figure~\ref{f:mode-ratios}).
This could be partly due to the smaller number statistics in the `B' simulations.
It is also possible that the effects of the secular modes are less pronounced because the hot classical test particles have higher inclinations than the Plutino test particles; perhaps because these test particles have to be scattered out further than the Plutino test particles, their inclinations are more affected by these random scattering events than by secular effects.

For the Plutino population at least, the correlation between the $f_8$ mode amplitude and median inclinations (as well as inclination distribution width) in our simulations is much clearer than any trend with migration timescale.
We checked for similar trends with the $f_7$ mode amplitude, but none was apparent.

A link between the $f_8$ mode amplitude and inclination excitation in the Plutino population provides a plausible explanation for the discrepancy between our simulation results and those of \citet{Nesvorny:2015}.
If Neptune's inclination is damped during planetary migration, as was done in \citet{Nesvorny:2015}'s simulations, this would also damp the amplitude of the $f_8$ secular mode, reducing the secularly forced inclinations of test particles in the outer solar system.
This could be particularly important for test particles passing through the $\nu_{18}$ secular resonance, where maximum forced inclinations due to the $f_8$ occur. 
In the current solar system, the location of the $\nu_{18}$ resonance is just beyond the 3:2 resonance, and based on the linear secular calculations this is also the case during all but the very beginning of our migration simulations.
So it is likely that many test particles that are scattered out past the 3:2 and are then picked up in the resonance as Neptune migrates experience some inclination excitation due to this resonance with the $f_8$ mode, as shown in Figure~\ref{f:secular-inclination-excitation}. 
Even particles that more directly stick to the 3:2 are likely to have spent some time at nearby semimajor axes and could thus be influenced.
We visually inspected the semimajor axis and inclination evolution of 50 test particles from each simulation that ended up in our final Plutino populations to estimate the relative frequency of secular inclination excitation associated with the $\nu_{18}$ resonance. 
In all but the $\tau_a=5$~Myr `B' simulation, approximately 50-70\% of the test particles showed significant changes in inclination while at constant semimajor axes just beyond the 3:2 resonance; in the $\tau_a=5$~Myr `B' simulation, which resulted in the fewest high inclination Plutinos, this fraction was $\sim30\%$.
For comparison, the percentage of test particles showing evidence of inclination excitation due to close encounters with Neptune (i.e., discrete changes in inclination correlated with discrete changes in semimajor axis) was also $\sim50-70\%$. 
The changes in inclination at constant semimajor axes near the estimated location of the $\nu_{18}$ resonance were smooth and in many cases roughly sinusoidal, which is consistent with the change being driven by secular forcing; some test particles exhibited very rapid inclination increases, consistent with being close to the center of the secular resonance.
We note that for test particles undergoing sinusoidal inclination changes, the net change in average inclination depends on the phase of the inclination cycle upon being captured into the 3:2 resonance, and some test particles had a net decrease in final inclination.
It is clear that the $f_8$ mode can contribute significantly to inclination excitation, and the relative importance of scattering events and secular evolution in the excitation of inclinations depends on the mode strength.
A strong or weak $f_8$ mode during migration could either enhance or reduce the
inclination excitation these test particles experience before ending up in the final Plutino population.
We also note that the mode amplitudes could affect the so-called `Kozai' resonance within the 3:2 mean motion resonance.
This resonance is characterized by the libration of  a Plutino's argument of perihelion and causes coupled opposite-phase variations in eccentricity and inclination  \citep[e.g.][]{Milani:1989}.
Kozai libration is likely to be affected by changes in the secular architecture of the planets because stationary values of the argument of perihelion correspond to a match between the precession rate of the longitude of perihelion and the regression rate of the longitude of ascending node. 
We leave an assessment of how the mode amplitudes affect the Kozai resonance and its corresponding inclination and eccentricity variations within the Plutino population for future work.

 The particular implementation of Neptune's inclination damping in the migration simulations could have different effects on Neptune's nodal rates in addition to influencing the mode amplitudes. (\cite{Nesvorny:2015} does not describe the specific scheme used in their simulations.)
In addition to affecting the amplitude of the $f_8$ mode by damping Neptune's inclination, the frequency of the mode itself could change, changing the locations of secular resonances that have a strong influence on particle inclinations.
Fixing the other giant planets' semimajor axes and migrating only Neptune would also affect the frequency and amplitude of the $f_8$ mode (and also secular resonance locations) in \cite{Nesvorny:2015}'s simulations compared to ours.
\citet{Nesvorny:2015} did perform some integrations where all four giant planets migrated and did not find evidence that this affected the correlation between inclinations and migration timescale in those simulations.  
However these additional simulations are not discussed in detail; it is possible that the initial and final conditions for these simulations differ from our simulations, resulting in different secular architectures.
Thus we find that a significant difference in the $f_8$ mode amplitude and frequency is the most plausible explanation for the dramatically different trends in our simulated Plutino populations compared to those in \cite{Nesvorny:2015}.

The giant planets' secular architecture should also affect the inclination distribution of the hot classical population, however it is possible that they are more influenced by encounters with Neptune than by secular inclination forcing.
\citet{Nesvorny:2015} found that it took of order a few tens of Myr for the inclinations of the population of Neptune-crossing test particles with semimajor axes in the classical belt range (the source population for the final implanted hot classical orbits) to become significantly dispersed as a result of encounters with Neptune (see his Figure~12); this led to the finding that hot classicals implanted late during the migration process would have higher inclinations and that slower migration would lead to higher inclinations in the final hot classical population.
In our simulations, we find a nearly identical trend to that found by \citet{Nesvorny:2015} in the inclination distribution of the hot classical source region with time,  but we do not find a strong trend between the final hot classical inclinations and migration speed.
The two longest migration timescale `A' simulations do have the broadest hot classical inclination distributions, but the $\tau_a=5$~Myr `B' simulation had the highest median inclination hot classical population of the `B' simulations.
Additionally, in the $\tau_a=5$~Myr `A' simulation, the final hot classical inclination distribution is broader than that of the source region after $\sim3$ e-folding timescales, after which very little migration occurs.  
This implies that while longer migration timescales can allow for more inclination excitation of the source population (due to more Neptune encounters, as suggested by \citealt{Nesvorny:2015}), there are additional mechanisms for exciting inclinations even with shorter migration timescales.
It seems plausible that secular excitation of inclinations can contribute to creating a broad hot classical inclination distribution.

In this work we have only investigated the inclination distributions that result from different migration scenarios in order to explain why strong trends between inclinations and migration speed seen in one set of simulations \citep{Nesvorny:2015} can be absent in others.
As discussed by \citet{Nesvorny:2015}, the observational estimates of inclination distributions for different Kuiper belt populations are often easier to obtain than other orbital parameter distributions or estimates of their total populations; thus, understanding how inclination distributions can be used to constrain the history of the outer solar system is of particular importance. 
However, many of the same points made above about the secular architecture of the planetary system also apply to the eccentricity distribution of Kuiper belt sub-populations.
Works such as \citet{Batygin:2011} and \cite{Dawson:2012} have shown that the low eccentricities of the cold classical Kuiper belt objects can be used to place constraints on Neptune's eccentricity and apsidal precession rate because these affect the amplitude of the secular variations in eccentricity experienced by small bodies in the Kuiper belt.
We have not investigated the evolution of the cold classical Kuiper belt in our simulations because there are a number of structures in their orbital distribution (such as the over-density of objects in the cold classical ``kernel'' at 44.5~au; \citealt{Petit:2011}) that are not reproduced by the kind of simplified migration scenarios investigated here and in \citet{Nesvorny:2015}. 
However, we note that our simulations are consistent with preserving the low eccentricities of the classical belt because we chose initial conditions such that Neptune's eccentricity remained small throughout migration.
The low inclinations of the cold classical belt are also likely to be preserved in our migration simulations, including those with enhanced $f_8$ mode strengths, because the location of the Kuiper belt's inclination secular resonance remains close to the 3:2 resonance throughout the simulation; we repeated our $tau_a=50$~Myr `A' simulation with a disk of test particles representing the cold classical belt and found that they did not experience significant inclination or eccentricity excitation. 

For the Plutino and hot classical populations, the location and strength of the $\nu_8$ eccentricity secular resonance associated with Neptune's dominant eccentricity secular mode will likely affect eccentricities similarly to how the $f_8$ mode and its $\nu_{18}$ secular resonance appears to affect inclinations.
Just as with inclination secular modes and amplitudes, the initial relative positions of the planets as well as the presence or absence of eccentricity damping will likely affect simulation outcomes and final secular architectures. 
 Previous studies have noted that the final giant planet secular architecture in planetary migration simulations is very sensitive to the planets' initial conditions and to whether planets encounter mean motion resonances or undergo mutual scattering events; this makes it difficult to perfectly reproduce the solar system's observed secular architecture in migration simulations \citep[e.g.,][]{Morbidelli:2009,Batygin:2010,Nesvorny:2012}.
Many previous investigations have focused on the eccentricity secular modes of Jupiter and Saturn because these modes have a strong influence on the stability of the terrestrial planets during giant planet migration \citep[e.g.,][]{Brasser:2009,Agnor:2012}.
For the Kuiper belt, both Neptune's eccentricity and inclination secular modes are important in sculpting the final orbital distributions.
Given how difficult it is to reproduce the current secular structure of the solar system at the end of migration simulations, the origin of Neptune's secular modes should be investigated in more detail in future work.

Finally, we note that the simplifying assumption of a massless test particle disk also affects the secular evolution of objects in these simulations. The eccentricities and inclinations of the massive pre-migration Kuiper belt could differ depending on how their self-gravity affects the evolution of both the planets and the KBOs \citep[see, .e.g.,][]{Hahn:2003,Reyes-Ruiz:2015}. 
While recent work by \cite{Fan:2017} shows that including self-gravity between the massive planetesimals in Nice-model like simulations does not appear to change the simulation outcomes for the planets or the overall inclination distribution for the resulting Kuiper belt particles, the details of the secular architecture and the locations of secular resonances could be affected; thus the inclinations of specific dynamical populations, such as the Plutino and hot classical populations, could be affected by the massive pre-migration planetesimal disk.

\section{Summary and Conclusions}\label{s:sum}

We have performed simplified simulations of giant planet migration to investigate the relationship between migration speed and inclination excitation in the Plutino and hot classical Kuiper belt populations.
As with all such investigations, these simulations do not represent the full, detailed dynamical history of the outer solar system, but instead allow us to better understand the relationship between migration speed and the inclination distributions of the Plutino and hot classical Kuiper belt populations.
We do not reproduce \citet{Nesvorny:2015}'s finding that slower migration speeds lead to more widely dispersed inclinations in these populations, instead finding no clear relationship between planet migration speed and inclination excitation for e-folding migration timescales of $\tau_a = $ 5, 10, 30, and 50 Myr.  
All of these migration timescales can yield inclination distributions of these populations that are broadly consistent with current observations.
For the Plutinos, we find that the degree of inclination excitation in our simulated populations is correlated with the amplitude of the $f_8$ inclination secular mode of the giant planets; this mode amplitude is sensitive to the simulated planets' initial conditions. 
Our simulated hot classical population shows only a very weak correlation between inclination excitation and the mode amplitude, perhaps indicating that scattering events are more important than secular effects for this population.
Our simulations are broadly similar to those of \citet{Nesvorny:2015}, but differences in the numerical implementation of planetary migration as well as  small differences in planetary initial conditions have a significant impact on the secular architecture of the simulated systems.
This plausibly leads to the different results regarding whether planet migration speed significantly controls inclination excitation in these Kuiper belt populations. 
The choice to migrate only Neptune or all four giant planets, or to damp or not damp planetary inclinations, affects the amplitudes of the inclination secular modes as well as the locations of secular resonances and their corresponding large forced inclinations in the Kuiper belt; the numerically simulated Kuiper belt orbital distributions can thus be very sensitive to the simplifications in numerical implementation of planet migration.
Our simulations indicate that planetary migration with e-folding timescales of 5 Myr, 10 myr, 30 myr and 50 myr, can all yield inclination dispersions similar to the observed Plutino and hot classical populations, with no correlation between the degree of inclination excitation and migration speed. 
Slow planetary migration is not necessarily required to explain the large inclinations.

We therefore conclude that constraints on the speed of planet migration must be sought in features other than the Kuiper belt inclination distribution. 
Some examples of potentially useful features have already been discussed in the literature.
These include \cite{Kaib:2016}'s suggestion that the orbital distribution of high-perihelion objects dropped from Neptune's 3:1 resonance could be strongly dependent on Neptune's migration speed.
The population ratios of objects captured in the different libration islands of Neptune's 2:1 mean motion resonance has also been found to potentially depend on migration speed \citep{Murray-Clay:2005}.
\cite{Murray-Clay:2011} suggest that the fraction of binaries in different subpopulations of the Kuiper belt might relate to migration speed. 
While the speed of migration might not be imprinted in the inclination distributions, other aspects of migration might be.
This work suggests that even small changes in the secular architecture of the planets can lead to significant changes in the inclination distributions of the Plutino population.
 Future studies to better understand how the planets' secular modes change during migration and how this affects Kuiper belt populations could lead to new constraints on the migration history of the outer solar system.

\vspace{-5pt}

\acknowledgements{
\noindent We thank the referee for helpful comments that improved the manuscript. We acknowledge funding from NASA (grants NNX14AG93G and 80NSSC19K0785) and NSF (grants AST-1312498 and AST-1824869).
An allocation of computer time from the UA Research Computing High Performance Computing (HPC) at the University of Arizona is gratefully acknowledged.}

\software{\sc{rebound}}

\listofchanges{}

\end{document}